\documentclass[11pt]{article}
\usepackage[left=3.1cm, right=3.2cm, top=3.5cm, bottom=3.25cm]{geometry}

\usepackage{cite}
\usepackage{amsmath,amssymb,amsfonts}
\usepackage{algorithmic}
\usepackage{graphicx}
\usepackage{textcomp}
\usepackage{xcolor}
\def\BibTeX{{\rm B\kern-.05em{\sc i\kern-.025em b}\kern-.08em
    T\kern-.1667em\lower.7ex\hbox{E}\kern-.125emX}}

\usepackage{tabularx}
\usepackage{tikz,pgfplots}
\usetikzlibrary{calc,patterns,angles,quotes}
\usepackage{mathtools}
\usepackage{xfrac}
\usepackage{siunitx}

\pgfplotsset{compat=1.18}

\graphicspath{{./figures/}}

\graphicspath{{./figures/}}

\newcounter{problemC}

\usepackage[capitalize,nameinlink,compress]{cleveref}
\crefname{appendix}{Appendix}{Appendices} 
\crefname{figure}{Figure}{Figures} 
\crefname{equation}{}{} 

\crefname{claim}{Claim}{Claims}
\crefname{example}{Example}{Examples}
\crefname{remark}{Remark}{Remarks}
\crefname{assumption}{Assumption}{Assumptions}
\crefname{fact}{Fact}{Facts}
\crefname{proposition}{Proposition}{Propositions}
\crefname{corollary}{Corollary}{Corollaries}
\crefname{lemma}{Lemma}{Lemmas}
\crefname{theorem}{Theorem}{Theorems}
\crefname{definition}{Definition}{Definitions}

\crefformat{problemC}{#2Problem~\ref{prob::#1}#3}
\crefmultiformat{problemC}{Problems~#2\cref{prob::#1}#3}{ and #2\cref{prob::#1}#3}{, #2\cref{prob::#1}#3}{ and #2\cref{prob::#1}#3}

\newcommand{\pcc}{\mathrm{pcc}}
\newcommand{\dc}{\mathrm{dc}}

\newcommand{\err}{\mathrm{err}}

\newcommand{\ac}{\mathrm{ac}}
\newcommand{\ff}{\mathrm{ff}}
\newcommand{\aaa}{\mathrm{al}}

\newcommand{\B}[1]{\boldsymbol{#1}}
\newcommand{\tran}{\top}

\newcommand{\dq}{\mathrm{dq}}
\newcommand{\dcom}{\mathrm{d}}
\newcommand{\qcom}{\mathrm{q}}

\newcommand{\ab}[1]{\alpha\beta}
\newcommand{\mpppc}{MP$^3$C}

\newcommand{\rf}{\mathrm{ref}}

\newcommand{\com}{\mathrm{sum}}
\newcommand{\diff}{\mathrm{diff}}
\newcommand{\tot}{\mathrm{t}}
\newcommand{\conv}{\mathrm{conv}}

\newcommand{\expm}[1]{e^{#1}}

\allowdisplaybreaks

\DeclareSIUnit \pu {pu}
\DeclareSIUnit \voltampere {VA}

\usepackage[footnotesize]{subfigure}

\pgfplotsset{compat=1.18}

\title{\Large\bf Damping Wind Farm Resonances with Current Based\\ Model Predictive Pulse Pattern Control
} 

\author{{Orcun Karaca}\\
	\textit{ABB Corporate Research Center} \\
		Baden-Dättwil, Switzerland \\
		orcun.karaca@ch.abb.com
	\and
{Ioannis Tsoumas}\\
	\textit{ABB System Drives} \\
		Turgi, Switzerland \\
		ioannis.tsoumas@ch.abb.com
	\and
{Tinus Dorfling}\\
\textit{ABB Corporate Research Center} \\
		Baden-Dättwil, Switzerland \\
		martinus.dorfling@protonmail.com \and 
	{Ran Chen}\\
	\textit{D-MAVT, ETH Zürich} \\
		Zürich, Switzerland \\
		rachen@student.ethz.ch
	\and
	{Lennart Harnefors}\\
	\textit{ABB Corporate Research Center} \\
		Västerås, Sweden \\
		lennart.harnefors@se.abb.com
}
\date{June 2024}

\begin{document}

\maketitle

\begin{abstract}
It is well-established that a proportional current control gain emulates a resistor in the converter output impedance. Even though this resistance can provide additional damping to grid resonances, its effect for traditional linear current controllers is known to be rather limited. Moreover, for medium-voltage systems, high switching frequencies are not an option due to the high switching losses. To meet the harmonic standards, it is expedient to use optimized pulse patterns.
This further exacerbates the problems with the resistance of classical controllers, since an additional filtering would be required so that the current controller acts only on the fundamental component (and not on the ripple component). Such a design limits the damping effect not only in its amplitude but also in the frequency range where it is active. This paper shows that a high-bandwidth current-based model predictive pulse pattern controller can alleviate these limitations. The pulse pattern control approach can achieve a high gain even at low switching frequencies, while controlling directly the instantaneous currents (i.e., the fundamental component and the ripple together). With a fast implementation cycle, the frequency range where this damping effect is active can be further extended. Numerical studies showcase these benefits for a multi-phase medium-voltage wind power conversion system.\\

\end{abstract}

\section{Introduction}

Offshore wind farms suffer from resonances created by the feeder cables and transformers. These resonances can lead to a high amplification of harmonics, often requiring expensive filters to be installed to sufficiently damp them~\cite{larsen2021managing}. A cost-effective solution here would be the one taking advantage of the converter current controller. The proportional gain of the traditional current controllers of grid-connected converters acts as an equivalent resistor at the converter output and thus can increase the converter effective resistance and enhance the system damping~\cite{harnefors2007input}. Nevertheless, control delays limit the damping effect to a rather low frequency range. At higher frequencies, the increasing phase shift results instead in a negative resistance contribution from the proportional part~\cite{sun2023high}. \looseness=-1

In medium-voltage applications, high switching frequencies are not an option due to the high switching losses, and this brings in an additional limitation to the effective resistance of the traditional controllers. Specifically, to meet the harmonic requirements of grid standards while also operating at a low switching frequency, it is desirable to use optimized pulse patterns (OPPs). OPPs are offline-calculated switching sequences that yield optimal harmonic distortions for a given pulse number (i.e., the number of pulses per phase within one fundamental period), see~\cite{patel1973generalized,buja1977optimal,rathore2012generalized,birth2019generalized} and the references therein. They allow incorporating constraints on specific harmonics.
In return, OPPs relinquish having a fixed modulation interval. Thus, regularly-spaced instants where the fundamental component can be conveniently sampled does not exist.
This results in both the fundamental component and the ripple being sampled, which complicates the control design. A typical solution for traditional linear controllers with a subsequent OPP modulator is to employ low pass filters, which would not only increase the total phase shift, but also limit the proportional gain. This leads to a reduction in the control-induced damping of resonances, and also a negative resistance at frequencies of only hundreds of Hz, amplifying the wind farm resonances present at those frequencies.\looseness=-1

Model predictive pulse pattern control (\mpppc)~\cite{geyer2011model} solves this problem, since it is designed to control the instantaneous quantities, i.e., both the fundamental component and the ripple, around their steady-state trajectories. Therefore, little-to-none low pass filtering is required. Though \cite{geyer2011model} has presented this controller for flux control, under certain modifications, it can also be adapted for current control.
A more recent work~in~\cite{dorfling2022generalized1,dorfling2022generalized2} has further generalized this method to higher-order systems typically found in grid-connected converter applications. Another work in~\cite{rosado2023selective} has experimentally validated that the \mpppc\ method can fulfill harmonic requirements while exhibiting a fast dynamic response.
The main motivation of this paper is to show that the current-based \mpppc\ method can alleviate the aforementioned limitations on the effective resistance of traditional controllers in medium-voltage applications, increasing both the amplitude and the frequency range of the control-induced damping.   \looseness=-1

Contributions of this paper are as follows. A linearized model is derived for the converter effective impedance of the current-based \mpppc\ method. Numerical case studies show that this model can closely approximate the underlying nonlinear system for different OPPs, pulse numbers, and under different multi-level converter topologies.
The superior passivity property of the current-based \mpppc\ method observed in the numerical case studies is attributed to \textit{(i)}~the control of the instantaneous currents with hardly any low-pass filtering, \textit{(ii)}~the model predictive pulse pattern control strategy, \textit{(iii)}~the~fast implementation cycle.\looseness=-1

The paper is organized as follows. Section~\ref{sec:pre} brings in the preliminaries on the modeling of a dual-converter wind power conversion system, and the current-based \mpppc\ method. Section~\ref{sec:model} derives the linearized model that will be used in the numerical case studies to anticipate the converter effective resistance. Numerical case studies are presented and discussed in Section~\ref{sec:res}.\looseness=-1

\begin{figure}[t!]
	\centering
	\resizebox{0.9\linewidth}{!}{
		\input{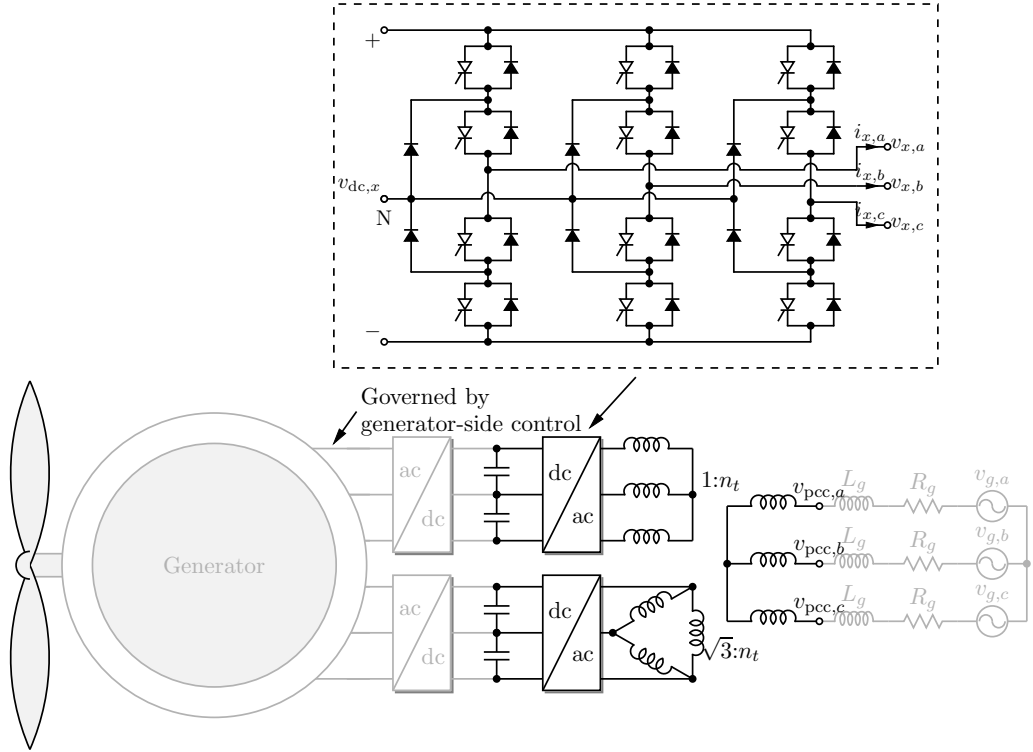}}
	\caption{Wind converter system comprising a dual conversion line connected to the grid through a transformer.}
	\label{fig::system}
\end{figure}

\section{Preliminaries}\label{sec:pre}

Consider a wind power conversion system that comprises a dual conversion line as depicted in the schematic in \cref{fig::system}. On the generator-side, the system is connected to a permanent magnet synchronous generator with two sets of three-phase windings. On the grid-side, the connection is via a transformer in a Wye-Delta-Wye configuration. Such a configuration has the well-established advantage of eliminating the $ 6n\pm 1\, (n=1, 3, \ldots) $ harmonics when a phase-shift of $ \pi/6\, $rad is applied between the two lines. \looseness=-1

The generator-side converters are approximated by controlled current sources that depend on the power produced by the generator.
The control unit of the generator-side converters is, among others, responsible for controlling the average value of the neutral-point potential of each conversion line, that is, balancing the upper and lower halves of the dc-link capacitors of each line.\looseness=-1

In the remainder of this section, we present the preliminaries on grid-side system modeling and control. For the inner current controller, which will be the current-based \mpppc\ method, the current references are assumed to have already been determined. 
The outer control loops of the grid-side converter provide these current references, for instance, to balance the total dc-link capacitor voltages of the two conversion lines, and to provide certain voltage support via reactive power injection. During low-voltage fault-ride-through, the current references are generally computed from a grid-code-defined formula~\cite{teodorescu_grid_2011}. In case of asymmetries, they would also include the negative sequence references, however, their incorporation into the inner control scheme is out-of-scope for this paper. \looseness=-1

The standard three-phase $abc$ frame is indicated as $\B{\xi}_{abc} = [\xi_a~\xi_b~\xi_c]^\tran$.
The $\alpha\beta$-reference frame (also known as the {stationary/orthogonal reference frame}) is denoted by $\B{\xi}_{\alpha\beta} = \xi_\alpha+j\xi_\beta$, and
obtained by the Clarke transformation, i.e.,  $$\B{\xi}_{\alpha\beta} = \B{T}_\mathrm{Clarke}\B{\xi}_{abc}=\frac{2}{3}\begin{bmatrix}
1 & -\frac{1}{2}+j\frac{\sqrt{3}}{2} & -\frac{1}{2}-j\frac{\sqrt{3}}{2}
\end{bmatrix}\B{\xi}_{abc}.$$  
The $\dq$-reference frame (also known as the synchronous rotating reference frame) is denoted by $\B{\xi}_{\dq} = \xi_\dcom+j\xi_\qcom$, and $\B{\xi}_{\dq} = \expm{-j\phi_\pcc}\B{\xi}_{\alpha\beta}.$
The $\dcom$-axis of the rotating reference frame is aligned (i.e.,~synchronized) with the PCC voltage. Thus, the angle that the $\dcom$-axis makes with the $\alpha$-axis is given by $\phi_\pcc = \omega_1 t+\phi_\mathrm{ini}$, where $\omega_1=2\pi f_1$ is the fundamental frequency of the grid, and $\phi_\mathrm{ini}$ is the initial value of the angle. Typically, $\phi_\mathrm{pcc}$ is obtained from a synchronous-reference frame phase-locked loop (PLL). Whenever the subscript is dropped from a quantity, the $\alpha\beta$-reference frame is to be assumed,  $\B{\xi} = \B{\xi}_{\alpha\beta}$.\looseness=-1

\subsection{System modeling}

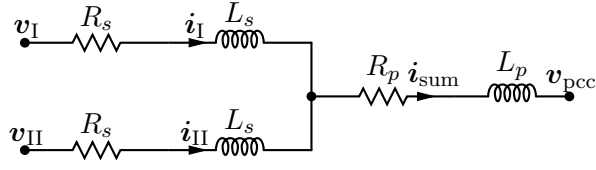
\begin{figure}[t]
	\centering
	\begin{tikzpicture}[scale=2.54]%
\ifx\dpiclw\undefined\newdimen\dpiclw\fi
\global\def\dpicdraw{\draw[line width=\dpiclw]}
\global\def\dpicstop{;}
\dpiclw=0.8bp
\dpiclw=0.8bp
\dpicdraw[fill=black](0,0) circle (0.007874in)\dpicstop
\draw (0.01,0) node[above=-2bp]{$\B{v}_\mathrm{I}$};
\dpicdraw[fill=black](0,-0.5625) circle (0.007874in)\dpicstop
\draw (0.01,-0.5625) node[above=-2bp]{$\B{v}_\mathrm{II}$};
\dpicdraw (0,0)
 --(0.25,0)
 --(0.270833,0.041667)
 --(0.3125,-0.041667)
 --(0.354167,0.041667)
 --(0.395833,-0.041667)
 --(0.4375,0.041667)
 --(0.479167,-0.041667)
 --(0.5,0)
 --(0.75,0)\dpicstop
\draw (0.375,0.041667) node[above=-2bp]{$ R_s$};
\dpicdraw (0.75,0)
 --(1,0)\dpicstop
\dpicdraw[line width=0.4bp](1,0) circle (0.00109in)\dpicstop
\dpicdraw (1,0)
 ..controls (1,0.034375) and (1.014625,0.0625)
 ..(1.0325,0.0625)
 ..controls (1.050375,0.0625) and (1.065,0.042813)
 ..(1.065,0.01875)
 ..controls (1.065,-0.005313) and (1.05825,-0.025)
 ..(1.05,-0.025)
 ..controls (1.04175,-0.025) and (1.035,-0.005313)
 ..(1.035,0.01875)
 ..controls (1.035,0.042813) and (1.053,0.0625)
 ..(1.075,0.0625)
 ..controls (1.097,0.0625) and (1.115,0.042813)
 ..(1.115,0.01875)
 ..controls (1.115,-0.005313) and (1.10825,-0.025)
 ..(1.1,-0.025)
 ..controls (1.09175,-0.025) and (1.085,-0.005313)
 ..(1.085,0.01875)
 ..controls (1.085,0.042813) and (1.103,0.0625)
 ..(1.125,0.0625)
 ..controls (1.147,0.0625) and (1.165,0.042813)
 ..(1.165,0.01875)
 ..controls (1.165,-0.005313) and (1.15825,-0.025)
 ..(1.15,-0.025)
 ..controls (1.14175,-0.025) and (1.135,-0.005313)
 ..(1.135,0.01875)
 ..controls (1.135,0.042813) and (1.153,0.0625)
 ..(1.175,0.0625)
 ..controls (1.197,0.0625) and (1.215,0.042813)
 ..(1.215,0.01875)
 ..controls (1.215,-0.005313) and (1.20825,-0.025)
 ..(1.2,-0.025)
 ..controls (1.19175,-0.025) and (1.185,-0.005313)
 ..(1.185,0.01875)
 ..controls (1.185,0.042813) and (1.199625,0.0625)
 ..(1.2175,0.0625)
 ..controls (1.235375,0.0625) and (1.25,0.034375)
 ..(1.25,0)\dpicstop
\dpicdraw[line width=0.4bp](1.25,0) circle (0.00109in)\dpicstop
\dpicdraw (1.25,0)
 --(1.5,0)\dpicstop
\draw (1.125,0.0625) node[above=-2bp]{$ L_s$};
\filldraw (0.86,-0.02125)
 --(0.945,0)
 --(0.86,0.02125) --cycle\dpicstop
\dpicdraw (0.922094,0)
 --(0.86,0)\dpicstop
\draw (0.891047,0) node[above=-2bp]{$ \B{i}_\mathrm{I}$};
\dpicdraw[line width=0.4bp](1.5,0) circle (0.00109in)\dpicstop
\dpicdraw (0,-0.5625)
 --(0.25,-0.5625)
 --(0.270833,-0.520833)
 --(0.3125,-0.604167)
 --(0.354167,-0.520833)
 --(0.395833,-0.604167)
 --(0.4375,-0.520833)
 --(0.479167,-0.604167)
 --(0.5,-0.5625)
 --(0.75,-0.5625)\dpicstop
\draw (0.375,-0.520833) node[above=-2bp]{$ R_s$};
\dpicdraw (0.75,-0.5625)
 --(1,-0.5625)\dpicstop
\dpicdraw[line width=0.4bp](1,-0.5625) circle (0.00109in)\dpicstop
\dpicdraw (1,-0.5625)
 ..controls (1,-0.528125) and (1.014625,-0.5)
 ..(1.0325,-0.5)
 ..controls (1.050375,-0.5) and (1.065,-0.519687)
 ..(1.065,-0.54375)
 ..controls (1.065,-0.567813) and (1.05825,-0.5875)
 ..(1.05,-0.5875)
 ..controls (1.04175,-0.5875) and (1.035,-0.567813)
 ..(1.035,-0.54375)
 ..controls (1.035,-0.519687) and (1.053,-0.5)
 ..(1.075,-0.5)
 ..controls (1.097,-0.5) and (1.115,-0.519687)
 ..(1.115,-0.54375)
 ..controls (1.115,-0.567813) and (1.10825,-0.5875)
 ..(1.1,-0.5875)
 ..controls (1.09175,-0.5875) and (1.085,-0.567813)
 ..(1.085,-0.54375)
 ..controls (1.085,-0.519687) and (1.103,-0.5)
 ..(1.125,-0.5)
 ..controls (1.147,-0.5) and (1.165,-0.519687)
 ..(1.165,-0.54375)
 ..controls (1.165,-0.567813) and (1.15825,-0.5875)
 ..(1.15,-0.5875)
 ..controls (1.14175,-0.5875) and (1.135,-0.567813)
 ..(1.135,-0.54375)
 ..controls (1.135,-0.519687) and (1.153,-0.5)
 ..(1.175,-0.5)
 ..controls (1.197,-0.5) and (1.215,-0.519687)
 ..(1.215,-0.54375)
 ..controls (1.215,-0.567813) and (1.20825,-0.5875)
 ..(1.2,-0.5875)
 ..controls (1.19175,-0.5875) and (1.185,-0.567813)
 ..(1.185,-0.54375)
 ..controls (1.185,-0.519687) and (1.199625,-0.5)
 ..(1.2175,-0.5)
 ..controls (1.235375,-0.5) and (1.25,-0.528125)
 ..(1.25,-0.5625)\dpicstop
\dpicdraw[line width=0.4bp](1.25,-0.5625) circle (0.00109in)\dpicstop
\dpicdraw (1.25,-0.5625)
 --(1.5,-0.5625)\dpicstop
\draw (1.125,-0.5) node[above=-2bp]{$ L_s$};
\filldraw (0.86,-0.58375)
 --(0.945,-0.5625)
 --(0.86,-0.54125) --cycle\dpicstop
\dpicdraw (0.922094,-0.5625)
 --(0.86,-0.5625)\dpicstop
\draw (0.891047,-0.5625) node[above=-2bp]{$ \B{i}_\mathrm{II}$};
\dpicdraw[line width=0.4bp](1.5,-0.5625) circle (0.00109in)\dpicstop
\dpicdraw (1.5,0)
 --(1.5,-0.5625)\dpicstop
\dpicdraw (1.5,-0.28125)
 --(1.75,-0.28125)
 --(1.770833,-0.239583)
 --(1.8125,-0.322917)
 --(1.854167,-0.239583)
 --(1.895833,-0.322917)
 --(1.9375,-0.239583)
 --(1.979167,-0.322917)
 --(2,-0.28125)
 --(2.25,-0.28125)\dpicstop
\draw (1.875,-0.239583) node[above=-2bp]{$ R_p$};
\filldraw (2.055,-0.3025)
 --(2.14,-0.28125)
 --(2.055,-0.26) --cycle\dpicstop
\dpicdraw (2.117094,-0.28125)
 --(2.055,-0.28125)\dpicstop
\draw (2.086047,-0.28125) node[left=-4bp, above=0bp]{$ \B{i}_\mathrm{sum}$};
\dpicdraw (2.25,-0.28125)
 --(2.425,-0.28125)\dpicstop
\dpicdraw[line width=0.4bp](2.425,-0.28125) circle (0.00109in)\dpicstop
\dpicdraw (2.425,-0.28125)
 ..controls (2.425,-0.246875) and (2.439625,-0.21875)
 ..(2.4575,-0.21875)
 ..controls (2.475375,-0.21875) and (2.49,-0.238438)
 ..(2.49,-0.2625)
 ..controls (2.49,-0.286563) and (2.48325,-0.30625)
 ..(2.475,-0.30625)
 ..controls (2.46675,-0.30625) and (2.46,-0.286563)
 ..(2.46,-0.2625)
 ..controls (2.46,-0.238438) and (2.478,-0.21875)
 ..(2.5,-0.21875)
 ..controls (2.522,-0.21875) and (2.54,-0.238438)
 ..(2.54,-0.2625)
 ..controls (2.54,-0.286563) and (2.53325,-0.30625)
 ..(2.525,-0.30625)
 ..controls (2.51675,-0.30625) and (2.51,-0.286563)
 ..(2.51,-0.2625)
 ..controls (2.51,-0.238438) and (2.528,-0.21875)
 ..(2.55,-0.21875)
 ..controls (2.572,-0.21875) and (2.59,-0.238438)
 ..(2.59,-0.2625)
 ..controls (2.59,-0.286563) and (2.58325,-0.30625)
 ..(2.575,-0.30625)
 ..controls (2.56675,-0.30625) and (2.56,-0.286563)
 ..(2.56,-0.2625)
 ..controls (2.56,-0.238438) and (2.578,-0.21875)
 ..(2.6,-0.21875)
 ..controls (2.622,-0.21875) and (2.64,-0.238438)
 ..(2.64,-0.2625)
 ..controls (2.64,-0.286563) and (2.63325,-0.30625)
 ..(2.625,-0.30625)
 ..controls (2.61675,-0.30625) and (2.61,-0.286563)
 ..(2.61,-0.2625)
 ..controls (2.61,-0.238438) and (2.624625,-0.21875)
 ..(2.6425,-0.21875)
 ..controls (2.660375,-0.21875) and (2.675,-0.246875)
 ..(2.675,-0.28125)\dpicstop
\dpicdraw[line width=0.4bp](2.675,-0.28125) circle (0.00109in)\dpicstop
\dpicdraw (2.675,-0.28125)
 --(2.85,-0.28125)\dpicstop
\draw (2.55,-0.21875) node[above=-2bp]{$ L_p$};
\definecolor{lcspec}{rgb}{0.10000,0.70000,0.70000}%
\color[rgb]{0.10000,0.70000,0.70000}%
\global\let\dpiclidraw=\dpicdraw\global\let\dpicfidraw=\filldraw1
\definecolor{fcspec}{rgb}{0.10000,0.70000,0.70000}%

\color{black}\global\let\dpicdraw=\dpiclidraw%
\global\let\filldraw=\dpicfidraw
\dpicdraw[fill=black](2.85,-0.28125) circle (0.007874in)\dpicstop
\draw (2.86,-0.28125) node[above=-2bp]{$\B{v}_{\pcc}$};
\dpicdraw[fill=black](1.5,-0.28125) circle (0.007874in)\dpicstop
\end{tikzpicture}%
	\caption{Equivalent circuit of a transformer with two secondary three-phase windings.}
	\label{fig:system-model}
\end{figure}
\begin{figure}[t!]
	\centering
	\resizebox{1.05\linewidth}{!}{
		\input{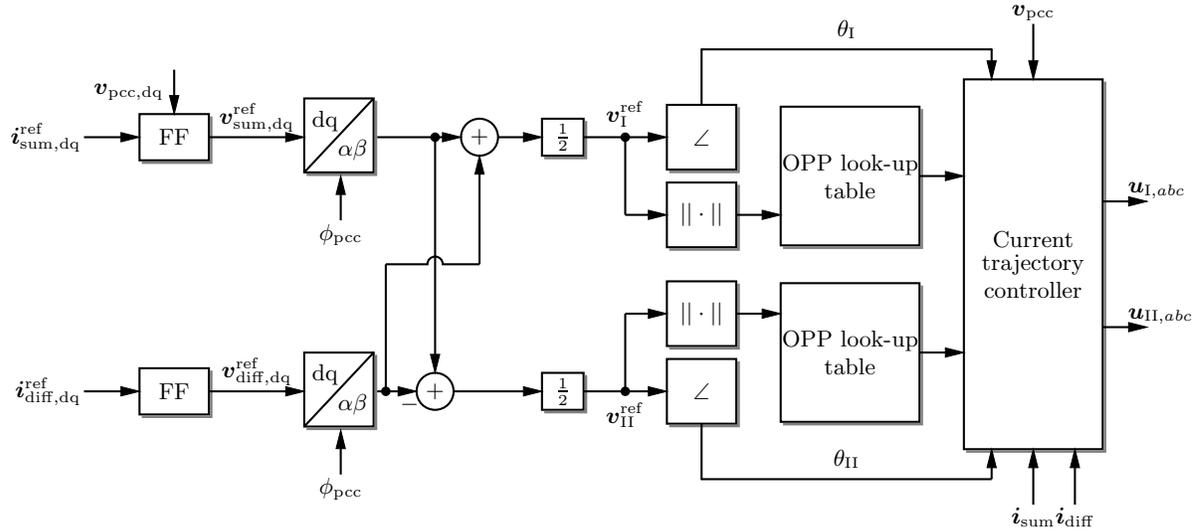}}
	\caption{A high-level block diagram for current-based \mpppc. }
	\label{fig::mp3c}
\end{figure}
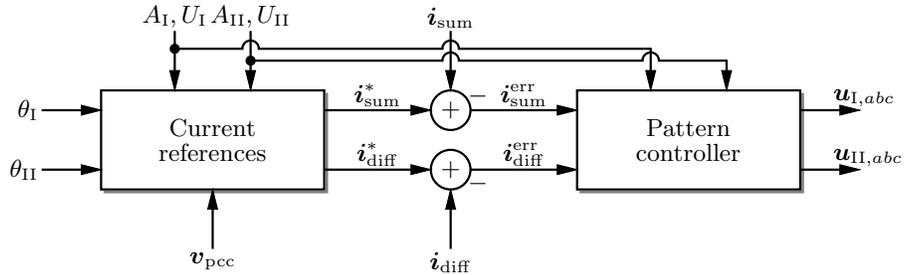
\begin{figure}[t!]
	\centering
	\resizebox{0.8\linewidth}{!}{
		\begin{tikzpicture}[scale=2.54]%
\ifx\dpiclw\undefined\newdimen\dpiclw\fi
\global\def\dpicdraw{\draw[line width=\dpiclw]}
\global\def\dpicstop{;}
\dpiclw=0.8bp
\dpiclw=0.8bp
\footnotesize
\dpicdraw (0,-0.25) rectangle (1.125,0.25)\dpicstop
\draw (0.5625,0) node{\shortstack{Current\\%
references}};
\dpicdraw[line width=1bp,draw=gray](0.0125,-0.2625)
 --(1.1375,-0.2625)
 --(1.1375,0.2375)\dpicstop
\dpicdraw (2.4,-0.25) rectangle (3.525,0.25)\dpicstop
\draw (2.9625,0) node{\shortstack{Pattern\\%
controller}};
\dpicdraw[line width=1bp,draw=gray](2.4125,-0.2625)
 --(3.5375,-0.2625)
 --(3.5375,0.2375)\dpicstop
\dpicdraw[fill=white](1.7625,0.15) circle (0.03937in)\dpicstop
\draw (1.7625,0.15) node{$+$};
\dpicdraw[fill=white](1.7625,-0.15) circle (0.03937in)\dpicstop
\draw (1.7625,-0.15) node{$+$};
\filldraw[line width=0bp](-0.1,0.125)
 --(0,0.15)
 --(-0.1,0.175) --cycle\dpicstop
\dpicdraw (-0.3,0.15)
 --(-0.022906,0.15)\dpicstop
\draw (-0.3,0.15) node[left=-2bp]{$\theta_\mathrm{I}$};
\filldraw[line width=0bp](-0.1,-0.175)
 --(0,-0.15)
 --(-0.1,-0.125) --cycle\dpicstop
\dpicdraw (-0.3,-0.15)
 --(-0.022906,-0.15)\dpicstop
\draw (-0.3,-0.15) node[left=-2bp]{$\theta_\mathrm{II}$};
\filldraw[line width=0bp](0.5875,-0.35)
 --(0.5625,-0.25)
 --(0.5375,-0.35) --cycle\dpicstop
\dpicdraw (0.5625,-0.55)
 --(0.5625,-0.272906)\dpicstop
\draw (0.5625,-0.55) node[below=-2bp]{$\B{v}_{\mathrm{pcc}}$};
\filldraw[line width=0bp](0.34625,0.35)
 --(0.37125,0.25)
 --(0.39625,0.35) --cycle\dpicstop
\dpicdraw (0.37125,0.55)
 --(0.37125,0.272906)\dpicstop
\draw (0.37125,0.55) node[above=-2bp]{$A_\mathrm{I},U_\mathrm{I}$};
\filldraw[line width=0bp](0.72875,0.35)
 --(0.75375,0.25)
 --(0.77875,0.35) --cycle\dpicstop
\dpicdraw (0.75375,0.55)
 --(0.75375,0.272906)\dpicstop
\draw (0.75375,0.55) node[above=-2bp]{$A_\mathrm{II},U_\mathrm{II}$};
\filldraw[line width=0bp](1.5625,0.125)
 --(1.6625,0.15)
 --(1.5625,0.175) --cycle\dpicstop
\dpicdraw (1.125,0.15)
 --(1.639594,0.15)\dpicstop
\draw (1.39375,0.15) node[above=-2bp]{$\B{i}^*_\com$};
\filldraw[line width=0bp](1.5625,-0.175)
 --(1.6625,-0.15)
 --(1.5625,-0.125) --cycle\dpicstop
\dpicdraw (1.125,-0.15)
 --(1.639594,-0.15)\dpicstop
\draw (1.39375,-0.15) node[above=-2bp]{$\B{i}^*_\diff$};
\filldraw[line width=0bp](2.3,0.125)
 --(2.4,0.15)
 --(2.3,0.175) --cycle\dpicstop
\dpicdraw (1.8625,0.15)
 --(2.377094,0.15)\dpicstop
\draw (2.13125,0.15) node[above=-2bp]{$\B{i}^\err_\com$};
\filldraw[line width=0bp](2.3,-0.175)
 --(2.4,-0.15)
 --(2.3,-0.125) --cycle\dpicstop
\dpicdraw (1.8625,-0.15)
 --(2.377094,-0.15)\dpicstop
\draw (2.13125,-0.15) node[above=-2bp]{$\B{i}^\err_\diff$};
\filldraw[line width=0bp](1.7375,0.35)
 --(1.7625,0.25)
 --(1.7875,0.35) --cycle\dpicstop
\dpicdraw (1.7625,0.55)
 --(1.7625,0.272906)\dpicstop
\draw (1.7625,0.55) node[above=-2bp]{$\B{i}_\com$};
\filldraw[line width=0bp](1.7875,-0.35)
 --(1.7625,-0.25)
 --(1.7375,-0.35) --cycle\dpicstop
\dpicdraw (1.7625,-0.55)
 --(1.7625,-0.272906)\dpicstop
\draw (1.7625,-0.55) node[below=-2bp]{$\B{i}_\diff$};
\filldraw[line width=0bp](3.725,0.125)
 --(3.825,0.15)
 --(3.725,0.175) --cycle\dpicstop
\dpicdraw (3.525,0.15)
 --(3.802094,0.15)\dpicstop
\draw (3.675,0.15) node[above right=-2bp]{$\B{u}_{\mathrm{I},abc}$};
\filldraw[line width=0bp](3.725,-0.175)
 --(3.825,-0.15)
 --(3.725,-0.125) --cycle\dpicstop
\dpicdraw (3.525,-0.15)
 --(3.802094,-0.15)\dpicstop
\draw (3.675,-0.15) node[above right=-2bp]{$\B{u}_{\mathrm{II},abc}$};
\dpicdraw[fill=black](0.37125,0.46) circle (0.007874in)\dpicstop
\dpicdraw (0.37125,0.46)
 --(0.717639,0.46)\dpicstop
\dpicdraw (0.712083,0.46)
 ..controls (0.712083,0.483012) and (0.730738,0.501667)
 ..(0.75375,0.501667)
 ..controls (0.776762,0.501667) and (0.795417,0.483012)
 ..(0.795417,0.46)\dpicstop
\dpicdraw (0.789861,0.46)
 --(1.726389,0.46)\dpicstop
\dpicdraw (1.720833,0.46)
 ..controls (1.720833,0.483012) and (1.739488,0.501667)
 ..(1.7625,0.501667)
 ..controls (1.785512,0.501667) and (1.804167,0.483012)
 ..(1.804167,0.46)\dpicstop
\dpicdraw (1.798611,0.46)
 --(2.77125,0.46)\dpicstop
\dpicdraw[line width=0.4bp](2.77125,0.46) circle (0.00109in)\dpicstop
\filldraw[line width=0bp](2.74625,0.35)
 --(2.77125,0.25)
 --(2.79625,0.35) --cycle\dpicstop
\dpicdraw (2.77125,0.46)
 --(2.77125,0.272906)\dpicstop
\dpicdraw[fill=black](0.75375,0.4) circle (0.007874in)\dpicstop
\dpicdraw (0.75375,0.4)
 --(1.726389,0.4)\dpicstop
\dpicdraw (1.720833,0.4)
 ..controls (1.720833,0.423012) and (1.739488,0.441667)
 ..(1.7625,0.441667)
 ..controls (1.785512,0.441667) and (1.804167,0.423012)
 ..(1.804167,0.4)\dpicstop
\dpicdraw (1.798611,0.4)
 --(2.735139,0.4)\dpicstop
\dpicdraw (2.729583,0.4)
 ..controls (2.729583,0.423012) and (2.748238,0.441667)
 ..(2.77125,0.441667)
 ..controls (2.794262,0.441667) and (2.812917,0.423012)
 ..(2.812917,0.4)\dpicstop
\dpicdraw (2.807361,0.4)
 --(3.15375,0.4)\dpicstop
\dpicdraw[line width=0.4bp](3.15375,0.4) circle (0.00109in)\dpicstop
\filldraw[line width=0bp](3.12875,0.35)
 --(3.15375,0.25)
 --(3.17875,0.35) --cycle\dpicstop
\dpicdraw (3.15375,0.4)
 --(3.15375,0.272906)\dpicstop
\draw (1.833211,0.220711) node[right=-2bp]{$-$};
\draw (1.833211,-0.220711) node[right=-2bp]{$-$};
\end{tikzpicture}%
}
	\caption{Current trajectory controller of current-based \mpppc.  }
	\label{fig::ic}
\end{figure}

A converter is denoted with $x \in \{\mathrm{I},\mathrm{II}\}$. Define $\B{v}_{x,abc} = [v_{x,a}~v_{x,b}~v_{x,c}]^\tran$ and $\B{u}_{x,abc} = [u_{x,a}~u_{x,b}~u_{x,c}]^\tran$ as the three-phase voltages and the switch positions of three-level neutral-point-clamped converter $x$, respectively, and $\B{v}_{x,abc} = \frac{v_{\dc,x}}{2}\B{u}_{x,abc}$, where $\frac{v_{\dc,x}}{2}$ is the half dc-link voltage. The voltage at the point of common coupling (PCC) is
denoted by $\B{v}_{\pcc,abc}$.

The output voltages of the converters, denoted by $\B{v}_{x}$, are described as
$\B{v}_{\mathrm{I}} = \B{T}_\mathrm{Clarke}\B{v}_{\mathrm{I},abc},$ $\B{v}_{\mathrm{II}} = D\B{T}_\mathrm{Clarke}\B{v}_{\mathrm{II},abc}$,
where $D = \expm{-j\tfrac{\pi}{6}}$
applies {a \ang{30}} clockwise rotation due to the delta configuration of the second conversion line. 

In Figure~\ref{fig:system-model}, the transformer is represented by its primary and secondary leakage inductances $L_p$ and $L_s$, and also the respective resistances $R_p$ and $R_s$,
The $\alpha\beta$-frame circuit dynamics for converters $\mathrm{I}$ and $\mathrm{II}$ are given by
\begin{equation}
	\begin{split}
		\B{v}_\mathrm{I} - [R_s +sL_s]\B{i}_\mathrm{I}- [R_p+sL_p](\B{i}_\mathrm{I}+\B{i}_\mathrm{II}) &= \B{v}_\pcc,\\
		\B{v}_\mathrm{II} - [R_s +sL_s]\B{i}_\mathrm{II}- [R_p+sL_p ](\B{i}_\mathrm{I}+\B{i}_\mathrm{II}) &= \B{v}_\pcc.
	\end{split}
\end{equation}
For dual conversion line systems, control design is often based on the sum and difference voltages and currents~\cite{karttunen2013decoupled}.  With $\B{\xi}_\mathrm{I}$ and $\B{\xi}_\mathrm{II}$ as the quantities of converters $\mathrm{I}$ and $\mathrm{II}$, these transformations are defined as 
$\B{\xi}_{\com} = \B{\xi}_\mathrm{I} + \B{\xi}_\mathrm{II},$ and 
$\B{\xi}_{\diff} = \B{\xi}_\mathrm{I} - \B{\xi}_\mathrm{II}$. 	
We can now reformulate the circuit equations above as
\begin{equation}\label{eq:dyn}
	\begin{split}
		\dfrac{\B{v}_\com}{2} -(R_\tot +sL_\tot)\B{i}_\com &= \B{v}_\pcc,\\
		{\B{v}_\diff} -(R_s+sL_s)\B{i}_\diff &= 0,
	\end{split}
\end{equation}
where $R_\tot=R_p+R_s/2$ and $L_\tot=L_p+L_s/2$. These equations can also be described in the (positive-sequence) $\dq$-reference frame as
\begin{equation}\label{eq:dyn_dq}
	\begin{split}
		\dfrac{\B{v}_{\com,\dq}}{2} -(R_\tot +sL_\tot+ j\omega_1L_\tot)\B{i}_{\com,\dq} &= \B{v}_{\pcc,\dq},\\
		{\B{v}_{\diff,\dq}} -(R_s+sL_s+ j\omega_1L_s)\B{i}_{\diff,\dq} &= 0.
	\end{split}
\end{equation}

\subsection{Current-based \mpppc}
In this section, we briefly describe the working principles of the \mpppc\ method in~\cite{geyer2011model}, with the sole focus being on its grid-side current control adaptation. Figure~\ref{fig::mp3c} presents the high-level block diagram that highlights the different components of the \mpppc\ method, whereas Figure~\ref{fig::ic} zooms into the current trajectory controller that includes the pulse pattern controller. 

Assume the fundamental current references $\B{i}_{\com,\dq}^\rf$ and $\B{i}_{\diff,\dq}^\rf$ are provided by the outer control loops. Invoking~\eqref{eq:dyn_dq}, the feedforward (FF) voltage references can be computed based on the steady state equations (i.e., $\frac{\mathrm{d}\B{i}_\dq}{\mathrm{d}t}=\B{0}$), 
\begin{equation}
	\begin{split}
		{\B{v}_{\com,\dq}^\rf}  &= 2\B{v}_{\pcc,\dq}+2(R_\tot+ j\omega_1L_\tot)\B{i}_{\com,\dq}^\rf,\\
		{\B{v}_{\diff,\dq}^\rf}  &= (R_s + j\omega_1L_s)\B{i}_{\diff,\dq}^\rf.
	\end{split}
\end{equation}
In a second step, these sum and difference voltage references are mapped to the converter voltage references in the $\alpha\beta$-reference frame to load the corresponding OPPs from the OPP look-up tables.

The current trajectory controller block first reads the instantaneous (i.e., including both the fundamental and the ripple) current references $\B{i}_\com^*$ and $\B{i}_\diff^*$ corresponding to the loaded OPPs---defined by a set of switching angles $A_x$ and a set of switching transitions $U_x$---and the operation point. If available, this part may utilize a slow time-scale grid impedance estimation, since the sum current ripple would depend on~it. Based on these references, current errors are computed: $\B{i}_\com^\err=\B{i}_\com^*-\B{i}_\com$ and $\B{i}_\diff^\err=\B{i}_\diff^*-\B{i}_\diff$. 

Given these errors, the pattern controller modifies the nominal OPPs provided by the pattern loaders, thus applying the patterns $\B{u}_{\mathrm{I},abc}$ and $\B{u}_{\mathrm{II},abc}$. The required pattern modifications are obtained by the \mpppc~method. This method is used in a receding horizon fashion, i.e., at each sampling instant, the corrections are computed, but only the current instances of the optimal inputs are applied, and the process is repeated at the next sampling instant. During steady-state operation, the grid-code harmonic requirements are met, since the nominal OPPs remain intact. To abide by the page limitations of the conference, we leave out a detailed description of the computations of these corrections. We refer the interested reader to~\cite[\S 12]{geyer_model_2016}.

As is described in~\cite[\S 12.4]{geyer_model_2016}, the pulse pattern controller itself is designed in a way (either as a deadbeat controller or as a quadratic program) to remove the error within a certain horizon by manipulating the switching instants. Because of this design principle, the \mpppc\ horizon can in fact give a good indication of the rise-time achieved by the overall scheme. 
For the case studies of this paper, the horizon will be specified as $T_{\mathrm{MP^3C}}=\frac{2}{f_1 p}$, where $p$ denotes the pulse number, that is, the number of pulses per phase within one fundamental period. This quantity will later be useful when approximating the gain of the current-based \mpppc~method to obtain a linear transfer function model. \looseness=-1

\section{Converter effective impedance modeling for current \mpppc}\label{sec:model}
\subsection{Impedance model and assumptions}
The current-based \mpppc\ method for the grid current, i.e., the sum current, will be modeled by the following linear controller in the $\alpha\beta$-reference frame;
\begin{equation}\label{eq:model}
	\begin{split}
		\dfrac{\B{v}_\com^\rf}{2}= K(s)&\left[  \B{i}_\com^*-H_\aaa(s)\B{i}_\com\right]\\&+H_\ff(s-j\omega_1)\left[R_\tot + j\omega_1L_\tot\right]\B{i}_{\com}^\rf\\&+ H_\pcc(s-j\omega_1)H_\aaa(s)\B{v}_{\pcc}.
	\end{split}
\end{equation}
The following remarks are in order for the assumptions of the model above.
\begin{enumerate}
	\item A linear approximation with transfer function $K(s)$ exists for the \mpppc\ method in the $\alpha\beta$-reference frame. Next subsection proposes one, which is later validated in simulations. 
	\item We have $\B{v}_\com^\rf=\B{v}_\com$, in other words, as opposed to the linear controllers, the \mpppc\ method manipulates the applied converter voltages directly via pattern control without relying on an external modulator. 
	\item The transfer function $H_\aaa(s)$ is an anti-aliasing filter designed based on the sampling rate of the \mpppc\ method.  
	\item The transfer function $H_\pcc(s-j\omega_1)$ is a low-pass filter. Notice that it has a $j\omega_1$ shift, since it is originally implemented in the $\dq$-reference frame~\cite{harnefors2007modeling}. Though not explicitly shown in its definition, this transfer function would also include the necessary zero-order holds and the delays for both the sampling of the PCC voltage and the control implementation. Similarly, $H_\ff(s-j\omega_1)$ includes the zero-order holds and delays related to the additional feedforward terms.
\end{enumerate}

With $\B{v}_\com^\rf=\B{v}_\com$, combining \eqref{eq:dyn} and \eqref{eq:model} we can derive
{\medmuskip=1mu
	\thickmuskip=1mu
	\begin{equation}
		\begin{split}
			\left[R_\tot +sL_\tot\right]\B{i}_\com+\B{v}_\pcc = K(s)&\left[\B{i}_\com^*-H_\aaa(s)\B{i}_\com\right]\\
			 &\hspace{-.642cm}+H_\ff(s-j\omega_1)\left[R_\tot + j\omega_1L_\tot\right]\B{i}_\com^\rf\\ &\hspace{-.642cm}+ H_\pcc(s-j\omega_1)H_\aaa(s)\B{v}_\pcc.
		\end{split}	
\end{equation}}

After regrouping and assuming that $\B{i}_\com^*$ and $\B{i}_\com^\rf$ are not varying as a function of $\B{v}_\pcc$ for the frequency range we are interested in\footnote{Such a coupling exist in the lower frequency range, e.g., within the bandwidth at which the dc-link controller operates, and it is in general asymmetric~\cite{harnefors2007input}.}, the converter effective impedance can be defined as follows
{\begin{equation}\label{eq:effimp}
	Z_\conv (s) = \dfrac{\B{v}_\pcc}{-\B{i}_\com}=\dfrac{K(s)H_\aaa(s) + R_\tot+sL_\tot}{1-H_\pcc(s-j\omega_1)H_\aaa(s)}.
\end{equation}}As a remark, in practice, the transformer resistance ${R_\tot}$ is not given by a constant, but it is a nondecreasing function of frequency due to skin effect. Studies with such transformer models are not included in this paper.

Finally, we conclude that the converter effective resistance at a given $\omega$ in the $\alpha\beta$-frame is simply the real part of the impedance:
$\Re\{Z_\conv (j\omega)\}$.
If the resistance is to be computed instead in the $\dq$-frame, a $j\omega_1$ shift is needed:
\begin{equation}
	\Re\{Z_{\conv,\dq} (j\omega)\} = \Re\{Z_{\conv} (j\omega+j\omega_1)\}.
\end{equation}

Next, we will derive the linear approximation $K(s)$ corresponding to the \mpppc\ method.
\subsection{A linear gain approximation}
For any first-order system, the well-known rise-time formula ties the 10\% to 90\% rise-time to the time constant:
\begin{equation}
	T_r = \tau \mathrm{ln}(9).
\end{equation}
It is straightforward to show that a purely inductive load with the inductance $L_\tot$ controlled with a simple proportional current controller results in the time constant $\tau = \frac{L_\tot}{K}$. Thus, we have
\begin{equation}
	K = \dfrac{L_\tot\mathrm{ln}(9)}{T_r}.
\end{equation}

For  the current-based \mpppc\ design used in this paper, we assume that the rise-time is equivalent to the \mpppc\ horizon, i.e., $T_r=T_{\mathrm{MP^3C}}=\frac{2}{f_1 p}.$
Thus, the gain is
\begin{equation}
	K = \dfrac{f_1 p}{2}\,L_\tot\mathrm{ln}(9).
\end{equation}

A computational delay is modeled by
\begin{equation}
	F_{\mathrm{delay}}^{T_d}(s)=e^{-sT_d},
\end{equation}
whereas a zero-order hold is modeled by
\begin{equation}
	F_{\mathrm{ZOH}}^{T_d}(s) =\dfrac{1-e^{-sT_d}}{sT_d}.
\end{equation}

Combining all the three components, the controller transfer function is given by
\begin{equation}
	K_{p,T_d} (s) = \dfrac{f_1 p}{2}\,L_\tot\mathrm{ln}(9)F_{\mathrm{ZOH}}^{T_d}(s) F_{\mathrm{delay}}^{T_d}(s).
\end{equation}
We obtain the converter effective resistance by plugging $K_{p,T_d} (s)$ into \eqref{eq:effimp}.
The numerical case studies will validate that this is indeed a meaningful approximation for the current-based \mpppc\ method.
\clearpage

\section{Numerical case studies}\label{sec:res}
In this section, we perform simulation studies to verify the theoretical converter effective impedance model for the current-based \mpppc\ method. The wind converter system is implemented in Matlab/Simulink. The rated values of the system are shown in Table~\ref{tab:rated values}. 
The relevant transformer parameters are $ R_\tot=\SI{3.1}{\ohm}$ and $ L_\tot=\SI{178}{\milli\henry}$.

The current-based~\mpppc\ method is implemented at the sampling rate $T_d=\SI{25}{\mu s}$. The cutoff frequencies of $H_\pcc$ and $H_\aaa$ are at $\SI{50}{Hz}$ and $\SI{15}{kHz}$, respectively. The impedance measurements are based on frequency sweeps, where the PCC voltage is perturbed at desired frequencies. Case studies at different sampling rates, the studies on negative frequencies, and the measurements close to the synchronous frequency are not included in this paper; but they are left out for future work.\looseness=-1

\begin{table}[t!]
	\caption{Rated values of the medium-voltage converter system.}
	\label{tab:rated values}
	\centering
	\begin{tabular}{lcl}
		\hline
		Parameter & Symbol & Value \\
		\hline
		Rated apparent power of a converter & $S_{\mathrm{R},\mathrm{conv}}$ & $\SI{7}{{\mega\voltampere}}$ \\ 
		Rated apparent power of the transformer & $S_{\mathrm{R},\mathrm{trafo}}$ & $\SI{14}{{\mega\voltampere}}$ \\ 
		Rated voltage at the primary (grid) & $V_{\mathrm{R},\mathrm{prim}}$ & $\SI{66}{\kilo\volt}$ \\
		Rated voltage at the secondary (converter) & $V_{\mathrm{R},\mathrm{sec}}$ & $\SI{3.1}{\kilo\volt}$ \\
		Rated angular frequency &  $\omega_1$ & $\SI[parse-numbers=false]{2 \pi 50}{\radian\per\second}$ \\
		\hline
	\end{tabular}
\end{table}

\subsection{A comparison at different operation points}
\begin{figure}[t]
	\centering
	\resizebox{0.7\linewidth}{!}{
		\includegraphics{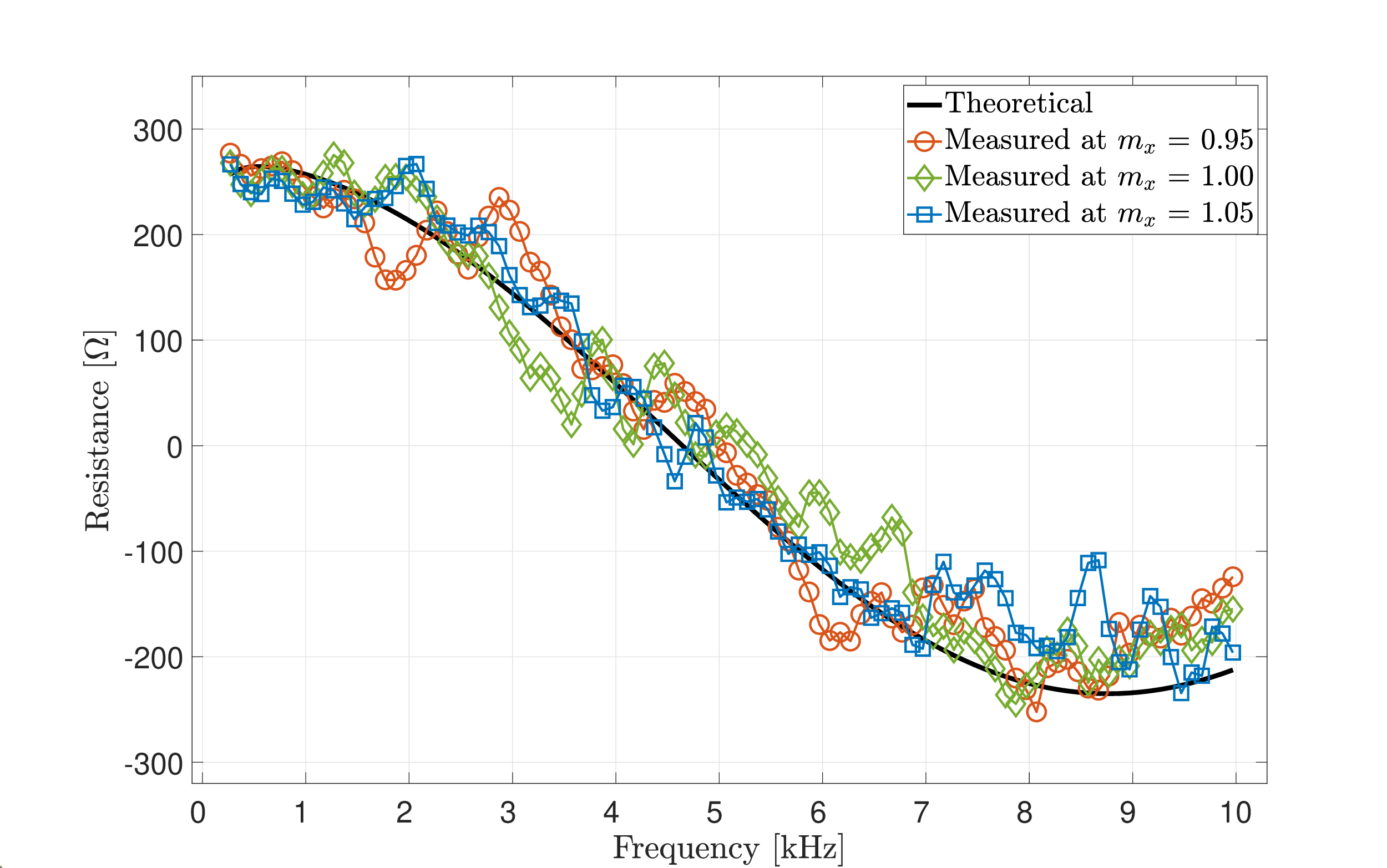}}
	\caption{Converter effective resistance for different operation points.}
	\label{fig::numA}
\end{figure}
We measure the converter effective resistance for different modulation indices: $m_x = \frac{2||\B{v}_{x}^{\rf}||}{v_{\dc,x}},$ using OPPs with the pulse number $p=14$, i.e., a device switching frequency of $\SI{350}{Hz}$.
Figure~\ref{fig::numA} shows the converter effective resistance and compares it with the theoretical model of the current-based~\mpppc. The theoretical model is accurate up to a mismatch of high-order terms. The measurements at different displacement angles have also turned out very similar to the theoretical curve.

The theoretical converter effective resistance crosses zero at $\SI{4.6}{{kHz}}$. As aforementioned, the resistance of the transformer would in practice increase with frequency. The crossover frequencies in Figure~\ref{fig::numA} are large enough such that the following negative resistance regions would simply be compensated by the transformer resistance. Note that a lower-bandwidth current measurement filtering required by linear controllers could move the crossover frequency to the left in a significant way. The observation that the negative resistance regions are compensated might then not necessarily hold.\looseness=-1

\subsection{A comparison with different OPPs but same pulse number}
\begin{figure}[t]
	\centering
	\resizebox{0.7\linewidth}{!}{
		\includegraphics{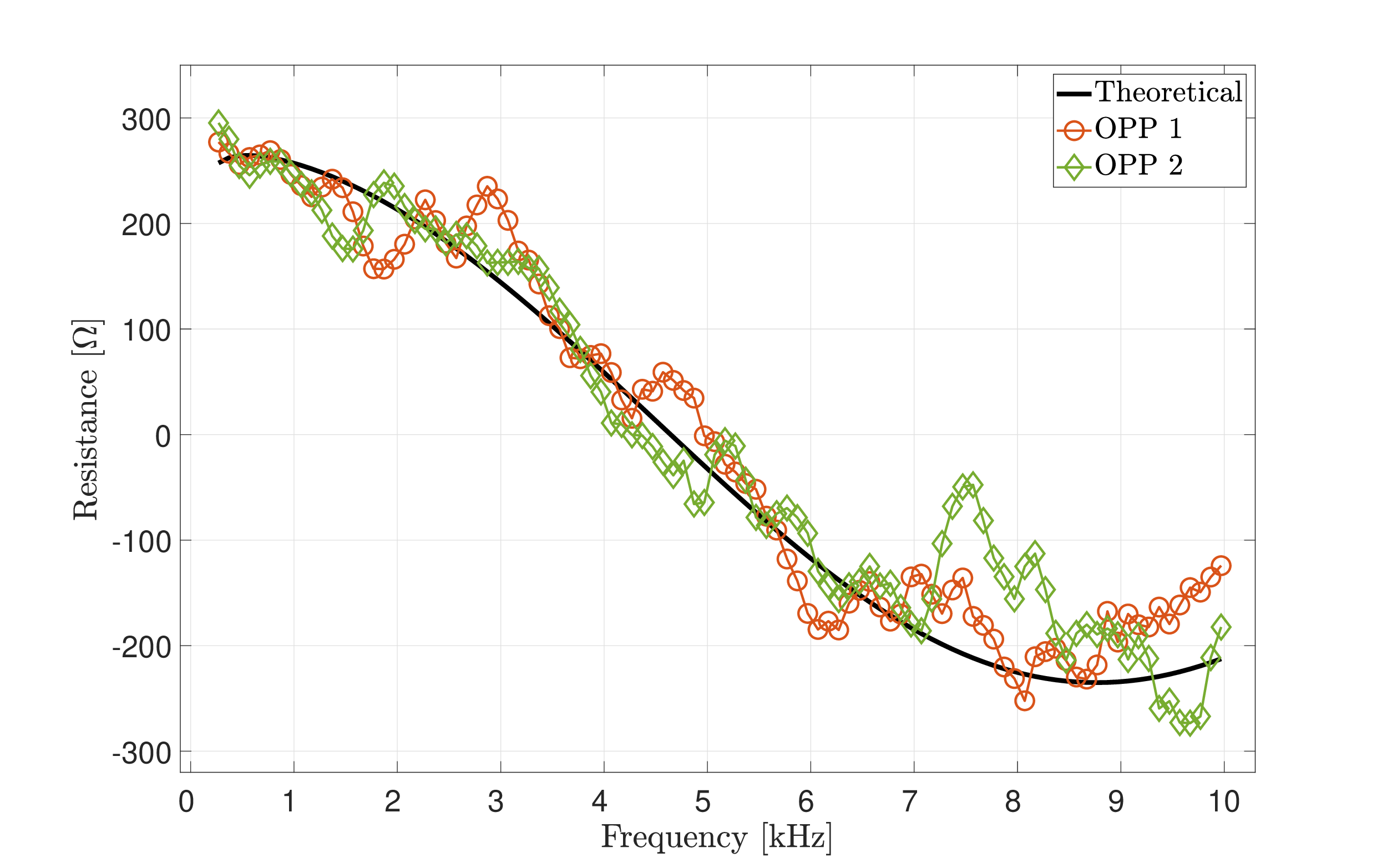}}
	\caption{Converter effective resistance for different OPPs with the same pulse number $p=14$.}
	\label{fig::numB}
\end{figure}

We pick $m_x = \SI{0.95}{\pu}$ and measure the converter effective resistance for different OPPs but with the same pulse number $p=14$. In other words, these OPPs result in the same device switching frequency but consist of different sets of switching angles. Figure~\ref{fig::numB} shows that these converter effective resistance measurements also follow closely the theoretical model. 

\subsection{A comparison with different pulse numbers}
\begin{figure}[t]
	\centering
	\resizebox{0.7\linewidth}{!}{
		\includegraphics{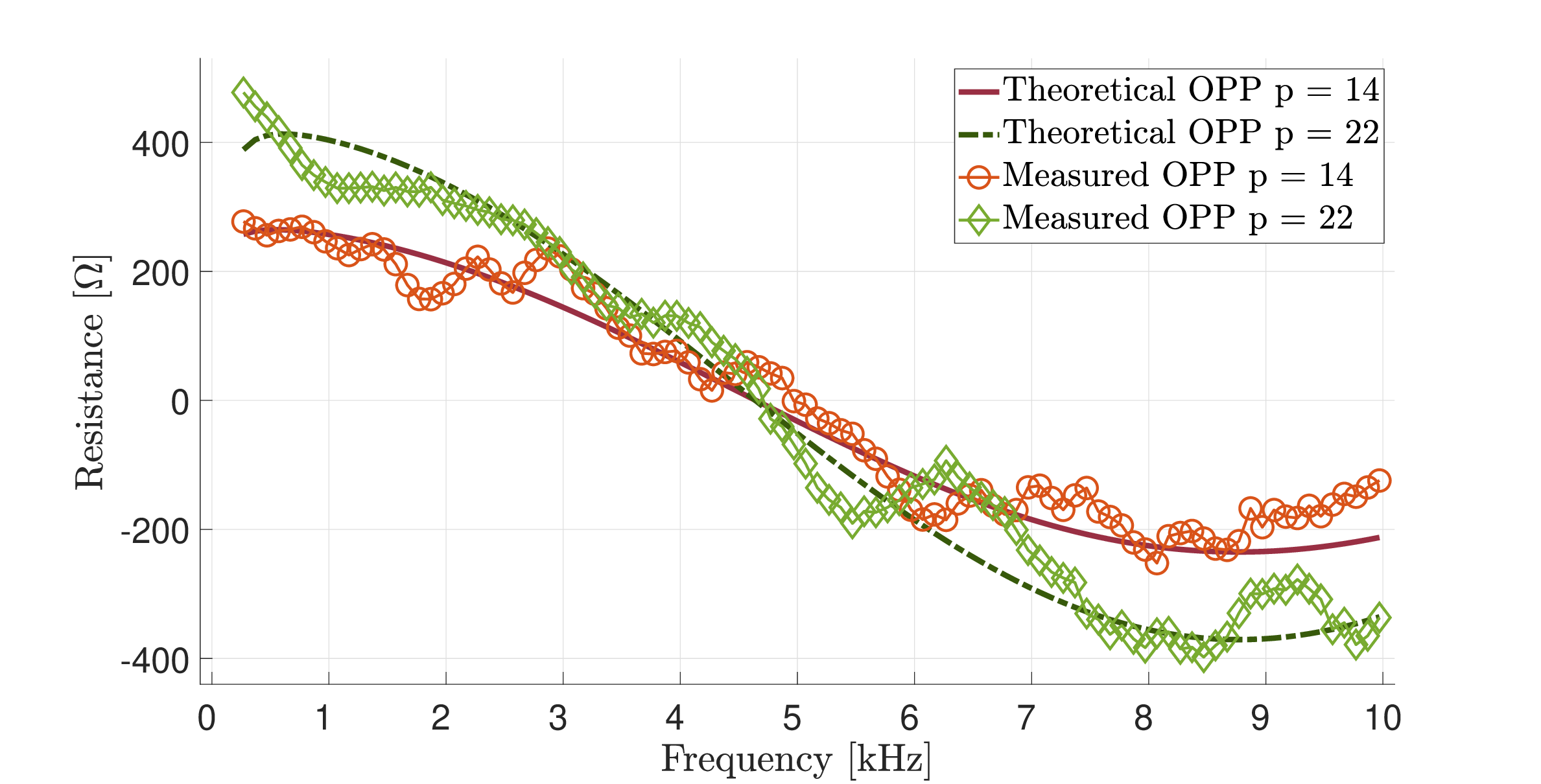}}
	\caption{Converter effective resistance for OPPs with different pulse numbers.}
	\label{fig::numC}
\end{figure}

We now measure the converter effective resistance for OPPs with different pulse numbers $p=14$ and $p=22$. Figure~\ref{fig::numC} shows each of these measurements follow their corresponding theoretical model. The \mpppc~ horizon specified in Section~\ref{sec:pre} implies that the gain increases linearly with the pulse number, and a similar increase can also be observed for the effective resistance.

\subsection{Increasing the effective resistance by a control modification}
\begin{figure}[t]
	\centering
	\resizebox{0.7\linewidth}{!}{
		\includegraphics{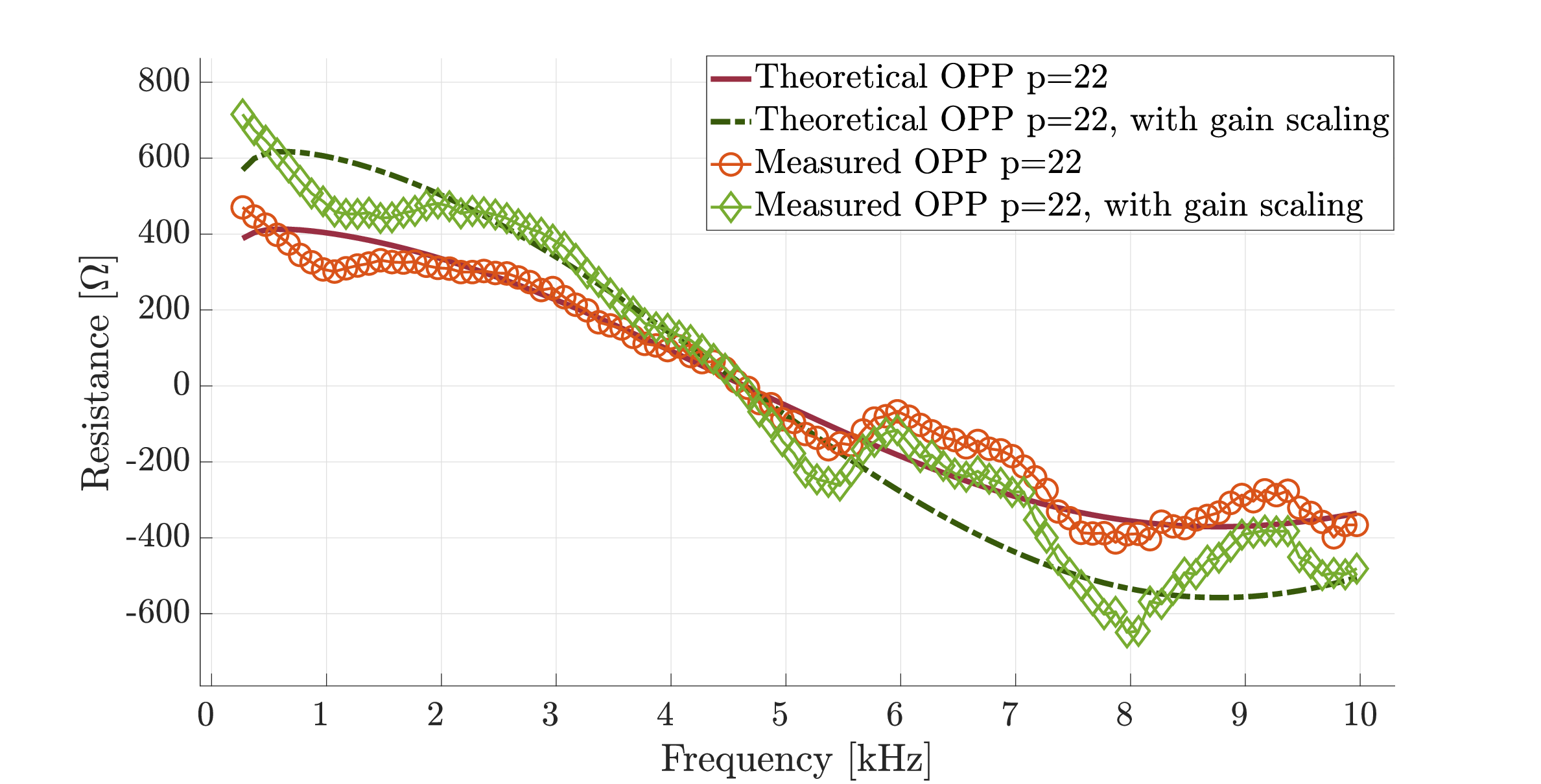}}
	\caption{The change in converter effective resistance when the error term is scaled up by a factor of $1.5$.}
	\label{fig::numD}
\end{figure}

For the case with $p=22$, we modify the current-based \mpppc~by scaling up the error by a factor of $3/2$. This modification reduces the rise time by a factor of $2/3$. Figure~\ref{fig::numD} illustrates the resulting converter effective resistance and compares it with its corresponding theoretical curve. 

\subsection{Multi-level converter case}
\begin{figure}[t]
	\centering
	\resizebox{0.7\linewidth}{!}{
		\includegraphics{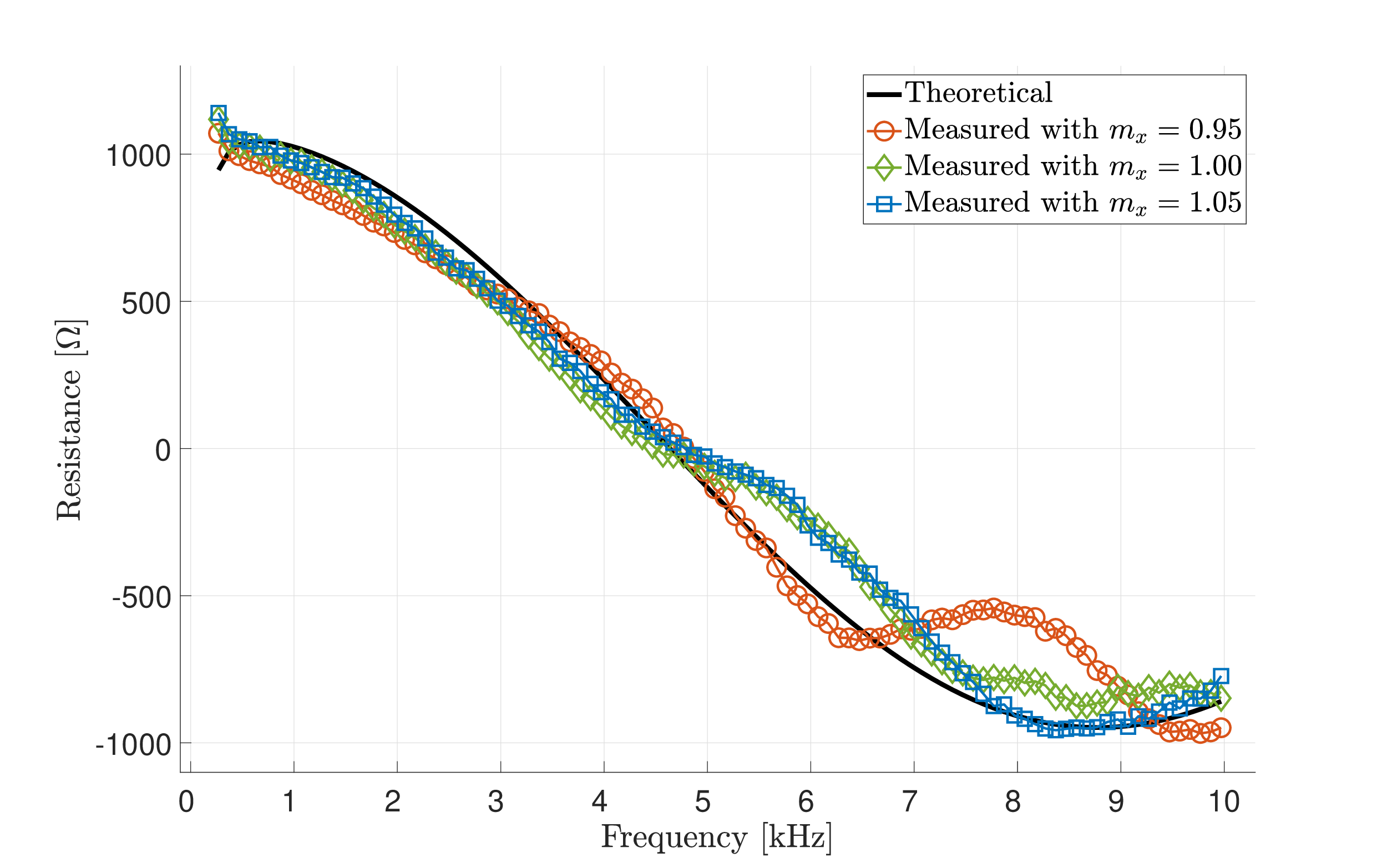}}
	\caption{Converter effective resistance of a $9$-level converter with an increased pulse number at $p=56$.}
	\label{fig::numE}
\end{figure}

We show that the results in the previous subsections for the 3-level converter hold also for other multi-level topologies. To this end, the current-based \mpppc\ is implemented instead with a 9-level converter topology with the pulse number $p=56$. Figure~\ref{fig::numE} illustrates the further increase in the effective resistance thanks to a high pulse number. The reduction in the ripple component appears to also reduce significantly the high-order terms in the impedance measurements.
\section{Conclusions}
A linearized model was derived for the converter effective impedance of the current-based \mpppc\ method. Numerical case studies verified that this model can anticipate the converter effective resistance for different OPPs, pulse numbers, and under different multi-level converter topologies.
The numerical case studies showcased the superior passivity of the current-based \mpppc\ method thanks to the control of the instantaneous currents without the need for filtering.

Future work could expand on this work with a larger set of case studies, including the negative sequences and the modeling of the outer and synchronization loops.

\section*{Acknowledgment}
The authors would like to thank Tobias Geyer at ABB System Drives, in Turgi, Switzerland for his input.
\bibliographystyle{IEEEtran}
\bibliography{report}

\end{document}